\documentclass[
 reprint,
 amsmath,
 amssymb,
 aps,
 prl, %
 superscriptaddress
]{revtex4-2}
\usepackage[utf8]{inputenc}
\usepackage{graphicx}
\usepackage{hyperref}
\usepackage{gensymb}
\usepackage{siunitx}
\usepackage{threeparttable, tablefootnote}
\usepackage{booktabs}
\usepackage{csquotes}
\usepackage[export]{adjustbox}
\usepackage{placeins}
\usepackage{setspace}
\usepackage{subcaption}
\hypersetup{breaklinks=true,
colorlinks=true,
linkcolor=blue,
citecolor=blue,
filecolor=magenta,
urlcolor=RoyalPurple}
\DeclareCaptionJustification{myjust}{\fulljustify}

\makeatletter
\providecommand\fulljustify{%
  \let\\\@centercr
  \leftskip\z@%
  \rightskip\z@%
  \parfillskip\z@\@plus 1fill\relax%
}
\usepackage[usenames,dvipsnames]{color}
\usepackage{cancel}

\newcommand{\iso}[2]{\ensuremath{^{#2}\mathrm{#1}}}
\newcommand{\ion}[2]{\ensuremath{^{#2}\mathrm{#1}^+}}
\newcommand{\gstransition}[0]{\ensuremath{\mathrm{S}_{1/2} (F=1, m_F = 0) \leftrightarrow~\mathrm{S}_{1/2} (F=0, m_{F} = 0)\ }}
\newcommand{\reportednumber}[1]{\textcolor{Black}{#1}}  
\newcommand{\systemnumber}[1]{\textcolor{Black}{#1}}  

\DeclareSIUnit[quantity-product = { }]
\year{\text{~yr}}
\DeclareSIUnit[quantity-product = { }]
\mbar{\text{~mbar}}
\DeclareSIUnit[quantity-product = { }]
\gauss{\text{G}}
\DeclareSIUnit[quantity-product = { }]
\Ci{\text{Ci}}

\newcommand{\gsAhfs}[0]{\SI{-27.684511052\pm0.000000005}{\GHz}}
\newcommand{\gsrferr}[0]{\SI{0.3}{\Hz}}
\newcommand{\gsqzerr}[0]{\SI{5}{\Hz}}
\newcommand{\gsstaterr}[0]{\SI{0.6}{\Hz}}
\newcommand{\gssyserr}[0]{\SI{5}{\Hz}}

\newcommand{\psAhfs}[0]{\SI{-5.447\pm0.004}{\GHz}}

\newcommand{\pszeemanerr}[0]{\SI{1.6}{\MHz}}
\newcommand{\psstaterr}[0]{\SI{120}{\kHz}}
\newcommand{\psdrifterr}[0]{\SI{3.2}{\MHz}}
\newcommand{\pssyserr}[0]{\SI{4.0}{\MHz}}

\newcommand{\dsAhfs}[0]{\SI[uncertainty-mode=full]{-619.7\pm1.1}{\MHz}}
\newcommand{\dsplitting}[0]{\SI{1.239\pm0.002}{\GHz}}

\newcommand{\dszeemanerr}[0]{\SI{0.8}{\MHz}}
\newcommand{\dsstaterr}[0]{\SI{100}{\kHz}}
\newcommand{\dsdrifterr}[0]{\SI{0.7}{\MHz}}
\newcommand{\dssyserr}[0]{\SI{1.1}{\MHz}}

\newcommand{\esbfield}[0]{\SI{1.1}{\gauss}}
\newcommand{\gsbfield}[0]{\SI{2.183\pm0.003}{\gauss}}

\newcommand{\qzcoeff}[0]{\SI[uncertainty-mode=full]{142.3\pm1.0}{\Hz\per\gauss\squared}}
\newcommand{\spam}[0]{\SI{0.9951\pm0.0009}{}}
\newcommand{\spamdetecttime}[0]{\SI{1.7}{\ms}}
\newcommand{\spamdarkphotons}[0]{\SI{0.04}{}}
\newcommand{\spambrightphotons}[0]{\SI{11.6}{}}
\newcommand{\spamthreshold}[0]{\SI{2.5}{}}
\newcommand{\spampulse}[0]{\SI{235}{\micro\second}}
\newcommand{\epszero}[0]{\SI{0.0035\pm0.0008}{}}
\newcommand{\epsone}[0]{\SI{0.007\pm0.002}{}}
\newcommand{\qzcoefftheory}[0]{\SI{141}{\Hz\per\gauss\squared}}
\newcommand{\detectionefficiency}[0]{\SI{0.2241\pm0.0011}{\%}}
\newcommand{\sFzeropreptime}[0]{\SI{100}{\micro\second}}

\newcommand{\numramseys}[0]{\SI{111}{}}
\newcommand{\numrabis}[0]{\SI{10}{}}
\newcommand{\dstarkpowers}[0]{\SIrange{1.7}{11.0}{\micro\watt}}
\newcommand{\pstarkpowers}[0]{\SIrange{5.0}{15.0}{\micro\watt}}
\newcommand{\bbrupperlimit}[0]{\SI{10}{\micro\Hz}}
\newcommand{\maxbfielddrift}[0]{\SI{9}{\micro\gauss\per\minute}}
\newcommand{\directnuclearcalculationpct}[0]{\SI{30}{\percent}}
\newcommand{\cservenyprecision}[0]{\SI{0.8}{\percent}}
\newcommand{\skipnikovprecision}[0]{\SI{0.5}{\percent}}
\newcommand{\bacupperlimit}[0]{\SI{60}{\milli\gauss}}
\newcommand{\groundstatediscrepancyMHz}[0]{\SI{47\pm26}{\MHz}}
\newcommand{\groundstatediscrepancysigma}[0]{\SI{3.6}{}}
\begin{document}

\preprint{APS/123-QED}

\title{Laser Cooling and Hyperfine Measurements of Radium-225 Ions}

\newcommand{\UCSBAffiliation}[0]{Department of Physics, University of California, Santa Barbara, California 93106, USA}

\author{Roy A. Ready}
\email{roy.a.ready@gmail.com}
\affiliation{\UCSBAffiliation}
\author{Haoran Li}
\affiliation{\UCSBAffiliation}
\author{Spencer Kofford}
\affiliation{\UCSBAffiliation}
\author{Robert Kwapisz}
\affiliation{\UCSBAffiliation}
\author{Huaxu Dan}
\affiliation{\UCSBAffiliation}
\author{Akshay Sawhney}
\affiliation{\UCSBAffiliation}
\author{Mingyu Fan}
\affiliation{\UCSBAffiliation}
\author{Craig Holliman}
\affiliation{\UCSBAffiliation}
\author{Xiaoyang Shi}
\affiliation{\UCSBAffiliation}
\author{Luka Sever-Walter}
\affiliation{\UCSBAffiliation}
\author{A. N. Gaiser}
\affiliation{Department of Chemistry, Michigan State University, East Lansing, Michigan 48824, USA}
\affiliation{Facility for Rare Isotope Beams, Michigan State University, East Lansing, Michigan 48824, USA}
\author{J. R. Griswold}
\affiliation{Radioisotope Science and Technology Division, Oak Ridge National Laboratory, Oak Ridge, Tennessee 37830, USA}
\author{Andrew M. Jayich}
\affiliation{\UCSBAffiliation}

\date{\today}

\begin{abstract}
$^{225}$Ra$^+$ ions (nuclear spin $I=1/2$) have transitions that are first-order insensitive to magnetic field noise, which is advantageous for optical clocks and quantum information science. We report on laser cooling and trapping of $^{225}$Ra$^+$ ions and hyperfine splitting measurements of the ion's 7s $^2\mathrm{S}_{1/2}$, 7p $^2\mathrm{P}_{1/2}$, and 6d $^2\mathrm{D}_{3/2}$\ states.
We measured the ground state hyperfine constant, $A(\mathrm{S}_{1/2}) =$ \SI{-27.684511052\pm0.000000005}{\GHz}, and the quadratic Zeeman coefficient, $C_2 =$ \SI[uncertainty-mode=full]{142.3\pm1.0}{} $\mathrm{Hz\ G}^{-2}$, of the $^2\mathrm{S}_{1/2} (F=0, m_F = 0) \leftrightarrow~^2\mathrm{S}_{1/2} (F=1, m_{F} = 0)$ transition. Our result addresses a discrepancy in the literature for the ground state hyperfine splitting.
We measured the hyperfine constants of the $^2\mathrm{P}_{1/2}$ state, $A(\mathrm{P}_{1/2}) =$ \SI{-5.447\pm0.004}{\GHz}, and the $^2\mathrm{D}_{3/2}$ state, $A(\mathrm{D}_{3/2}) =$ \SI[uncertainty-mode=full]{-619.7\pm1.1}{\MHz}. We also performed state preparation and measurement using the ground state hyperfine levels and realized a fidelity of \SI{0.9951\pm0.0009}{}.

\end{abstract}

\maketitle

Atoms with magnetic field-insensitive transitions are appealing for quantum information science (QIS)~\cite{Crain2014} and precision measurements~\cite{Lewty2012, Pizzocaro2017}.
For alkali-like ions with a single unpaired electron and half-integer nuclear spin, there is a ground state qubit composed of two spin projection $m_F=0$ states separated by the hyperfine splitting.
These states are first-order insensitive to magnetic field noise and are well-suited to laser driven gates~\cite{Ballance2016, Boguslawski2023}.
For laser-coolable ions with nuclear spin $I=1/2$, the qubit is straightforward to initialize and control optically or with microwaves.
Such a hyperfine qubit is a feature of $^{171}$Yb$^+$~\cite{Balzer2006, Olmschenk2007} and $^{133}$Ba$^+$ ~\cite{Hucul2017}.
These qubit features also benefit optical clocks~\cite{Tamm2014, Huntemann2016}.
The \ion{Ra}{225} ion ($I=1/2$) is a good clock candidate \cite{Holliman2022}, due in part to its low charge to mass ratio and optical transitions at convenient wavelengths~\cite{Fan2019, West2019}.
The ion's large hyperfine splittings make it possible to operate an optical clock with only two infrared wavelengths (828~nm and 1079~nm)~\cite{Holliman2023}, which could be generated and controlled with integrated photonics to help enable a compact transportable system.

Here we report spectroscopy results with single trapped and laser cooled \ion{Ra}{225} ions.
We improve the precision of $^2\mathrm{S}_{1/2}$\ and $^2\mathrm{P}_{1/2}$\ hyperfine splittings and realize the first $^2\mathrm{D}_{3/2}$ hyperfine splitting measurement.
We also characterized state preparation and measurement (SPAM) of the hyperfine qubit using the \mbox{$\mathrm{S}_{1/2} \leftrightarrow \mathrm{P}_{1/2}$} and \mbox{$\mathrm{D}_{3/2} \leftrightarrow \mathrm{P}_{1/2}$} transitions, demonstrating straightforward, high fidelity control.

\begin{figure}
    \includegraphics[]{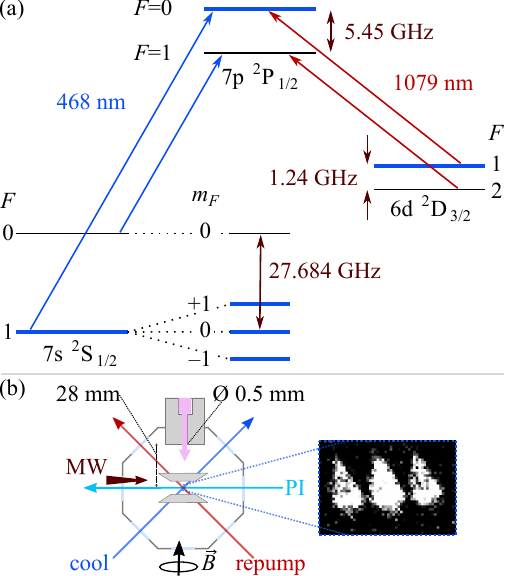}
    \captionsetup{justification=myjust,singlelinecheck=false}
    \caption{
(a)~Energy level diagram of the low-lying \ion{Ra}{225} hyperfine structure and $^2\mathrm{S}_{1/2}$ Zeeman sublevels, showing laser and microwave-driven (MW) transitions.
State detection lasers drive transitions between the bright states (blue) and are off-resonant from the dark states (black).
(b)~Experimental apparatus.
An oven emits radium atoms (pink arrow) that are photoionized by (PI) light at 483~nm and 450~nm.
Trapped ions, such as the three shown, are detected and laser cooled with light at 468~nm.
Our objective introduces aberrations to the image but does not affect our measurements.
Permanent magnets define the quantization axis ($\vec{B}$) and hyperfine transitions are driven by a microwave horn.
    }\label{fig:hfs_diagram}
\end{figure}

Radium is famous for its radioactivity and \iso{Ra}{225} is no exception with only a \systemnumber{\SI{15}{\day}} half-life.
Despite this apparent obstacle we have worked with laser cooled \ion{Ra}{225} ions in a sealed vacuum system since \mbox{June 6$^\text{th}$, 2023}.
The key to seamless operation is the oven design based on \mbox{Fan \textit{et al.}~\cite{Fan2023a}}.
We applied the same technique to thorium-229 (\systemnumber{\SI{7800}{\year}} half-life~\cite{Essex2018}), which continuously produces radium via nuclear decay.
The \iso{Ra}{225} source is a molecular beam epitaxy oven with \systemnumber{\mbox{\SI{8}{\micro\Ci}}} of \iso{Th}{229} deposited in a titanium crucible~\cite{Fan2023a}, which generates \systemnumber{\SI{3e11}{}} atoms over one \iso{Ra}{225} half-life.
The oven is heated to between \systemnumber{\SI{350}{\degreeCelsius} and \SI{500}{\degreeCelsius}} to effuse radium atoms, see Fig.~\ref{fig:hfs_diagram}\textcolor{blue}{b}.
Radium ions are loaded into the trap with two-stage photoionization: neutral radium is driven to the $^1\mathrm{P}_1$\ state with near-resonant 483~nm light and then ionized with 450~nm light.
Radium ions are consistently trapped within \systemnumber{\SI{15}{\min}} of the oven reaching the target temperature.
This long-term \iso{Ra}{225} source opens the door to working with \ion{Ra}{225} ions with high duty cycles in sealed systems.

To laser cool \ion{Ra}{225}, we drive optical transitions from hyperfine levels of the  $^2\mathrm{S}_{1/2}$ and $^2 \mathrm{D}_{3/2}$ states to the $^2\mathrm{P}_{1/2}$ state, see Fig.~\ref{fig:hfs_diagram}\textcolor{blue}{a}.
The ions are held in a linear Paul trap with characteristic radial and axial dimensions \mbox{\systemnumber{$r_0 =$~\SI{0.6}{\mm}}} and \mbox{\systemnumber{$z_0 =$~\SI{2.5}{\mm}}}~\cite{Fan2021}.
The radial electrodes are driven at \systemnumber{\SI{8.205}{\MHz}} and dc~offsets up to \systemnumber{\SI[retain-explicit-plus]{+200}{\milli\V} and \SI{-200}{\milli\V}} are applied to opposite electrode pairs to break the degeneracy between radial motional modes.
We can trap chains of ions, see Fig.~\ref{fig:hfs_diagram}\textcolor{blue}{b}, though measurements were performed with single ions.
Four lasers address all hyperfine levels in the $^2\mathrm{S}_{1/2}$ ground and  $^2\mathrm{D}_{3/2}$ metastable states, see Fig.~\ref{fig:hfs_diagram}\textcolor{blue}{a}.
The frequency and power of each laser is independently controlled with acousto-optic modulators set up in double-pass configurations.
Decays from the $^2\mathrm{P}_{1/2}$ state populate both the ground state and 
the metastable $^2\mathrm{D}_{3/2}$\ state~(\systemnumber{\SI{642\pm9}{\ms}}~lifetime~\cite{Li2025}).
The $^2\mathrm{D}_{3/2}$\ state hyperfine levels are repumped with 1079~nm light, which returns the ion to the ground state via the $\mathrm{D}_{3/2} \leftrightarrow \mathrm{P}_{1/2}$ transition.
Both 468~nm lasers and one repump laser are stabilized to an optical cavity with a finesse of \systemnumber{$\approx 1000$} and a laser linewidth below \systemnumber{\SI{2}{\mega\hertz}}.
The second repump laser is offset locked~\cite{Schuenemann1999} to its counterpart.
We use the first-order sidebands of fiber electro-optic modulators (EOMs) to scan the 468~nm and 1079~nm light across the hyperfine levels and also use the 468~nm EOM for state preparation, following \mbox{Olmschenk \textit{et al.}~\cite{Olmschenk2007}.}

\begin{figure*}
    \begin{subfigure}{0.252\textwidth}
        \includegraphics[scale=1]{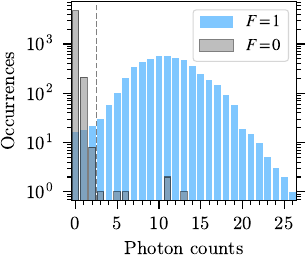}\vspace{-.5em}
        \caption{}
        \label{fig:spam}
    \end{subfigure}
~~~~
    \begin{subfigure}{0.282\textwidth}
        \includegraphics[scale=1]{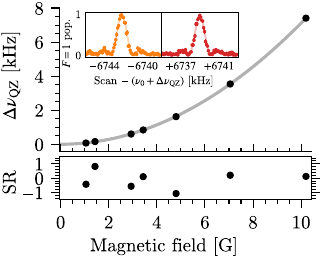}\vspace{-.5em}
        \caption{}
        \label{fig:qz-shift}
    \end{subfigure}\vspace{-.5em}
~~
    \begin{subfigure}{0.292\textwidth}
        \includegraphics[scale=1]{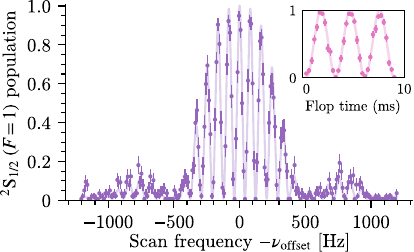}\vspace{-.5em}
        \caption{}
        \label{fig:ramsey-scan}
    \end{subfigure}\vspace{-.5em}
\hfill
    \captionsetup{justification=myjust,singlelinecheck=false}
    \caption{
    (a)~State preparation and measurement result for the ground state hyperfine transition.
    We measure an average of \reportednumber{\spamdarkphotons~photons} in the dark state and \reportednumber{\spambrightphotons~photons} in the bright state.
    (b)~The quadratic Zeeman shift for a range of magnetic field strengths, with studentized residuals (SR) from the quadratic fit used to extract $C_2$, in the bottom panel.
    We determine the magnetic field magnitude at the ion by driving  $\sigma^+$ (orange) and $\sigma^-$ (red) transitions from the \mbox{$m_F=\pm1$} Zeeman levels, shown for the \mbox{\SI{1.5}{\gauss}} data point.  
    (c)~Spectroscopy of the \gstransition transition, offset by $\nu_\mathrm{offset} = \systemnumber{\SI{27684511730}{\hertz}}$.
    A $\pi/2$ pulse time is determined from a Rabi flop, an example of which is shown (pink inset).
    }
\end{figure*}

Microwave spectroscopy of the $^2\mathrm{S}_{1/2}$ state is performed with a  horn positioned \systemnumber{\SI{10}{\cm}} from the trap center, see Fig.~\ref{fig:hfs_diagram}\textcolor{blue}{b}.
The microwaves are generated by mixing a fixed tone at \systemnumber{\SI{27.3}{\GHz}} with a tunable rf~source from a direct digital synthesizer that is frequency doubled to \systemnumber{\SI{385}{\MHz}}.

Each hyperfine splitting measurement consists of Doppler cooling, state preparation, driving the spectroscopy transition, and  state detection.
For \mbox{$^2\mathrm{S}_{1/2} (F=0)$} state preparation, we drive the \mbox{$\mathrm{S}_{1/2} (F=1) \leftrightarrow~\mathrm{P}_{1/2} (F=1)$}\ transition  for \systemnumber{\sFzeropreptime} while repumping both $\mathrm{D}_{3/2}$ hyperfine levels.
During state detection, the \mbox{$\mathrm{S}_{1/2} (F=1) \leftrightarrow~\mathrm{P}_{1/2} (F=0)$} transition is driven.
To prevent leakage to the $^2\mathrm{D}_{3/2}$\ (F=1) state, we repump population using the \mbox{$\mathrm{D}_{3/2} (F=1) \leftrightarrow~\mathrm{P}_{1/2} (F=0)$}\ transition.
The \mbox{$^2\mathrm{S}_{1/2}$} \mbox{$(F=1)$} and \mbox{$^2\mathrm{D}_{3/2} (F=1)$} states are referred to as bright states, shaded blue in Fig.~\ref{fig:hfs_diagram}\textcolor{blue}{a}.
Otherwise the ion is in one of the dark states, \mbox{$^2\mathrm{S}_{1/2} (F=0)$} or \mbox{$^2\mathrm{D}_{3/2} (F=2)$}, and light is not scattered as the dark states are far off resonance from the state detection lasers.
Scattered 468~nm photons are focused onto a photomultiplier tube (PMT) and camera by an objective.

We characterize the hyperfine qubit by measuring the SPAM fidelity.
After preparing the dark ground state, we transfer the population to the \mbox{$^2\mathrm{S}_{1/2} (F=1, m_F=0)$} bright state with a \systemnumber{\spampulse} resonant microwave pulse and perform state detection.
This is compared to state detection after only \mbox{$^2\mathrm{S}_{1/2} (F=0)$} dark state preparation (no microwave pulse).
From the SPAM measurement we find a \mbox{\reportednumber{\spamthreshold~photon}} threshold to discriminate between bright events (\reportednumber{\spambrightphotons~photons on average}) and dark events (\reportednumber{\spamdarkphotons~photons on average}) in a \reportednumber{\spamdetecttime} detection time, see Fig.~\ref{fig:spam}.
The threshold minimizes the average false detection fraction, given by \mbox{$1/2\left(\epsilon_0 + \epsilon_1\right)$}~\cite{Christensen2020}, where $\epsilon_0$\ ($\epsilon_1$) is the fraction of dark (bright) events falsely identified as bright (dark) events, and is related to the fidelity by \mbox{$\mathcal{F} = 1 - 1/2\left(\epsilon_0 + \epsilon_1\right)$}.
We determine the fidelity from an average of \systemnumber{13}~measurements, with a single representative measurement shown in Fig.~\ref{fig:spam}.
The measured average SPAM fidelity is \mbox{$\mathcal{F} =$ \reportednumber{\spam}}, with false identification fractions of \mbox{$\epsilon_0 = $ \reportednumber{\epszero}} and \mbox{$\epsilon_1 = $ \reportednumber{\epsone}}.
The result is limited by off-resonant scattering, PMT dark counts, and the measured (with statistical uncertainty) \reportednumber{\detectionefficiency} photon detection efficiency~\cite{Ramm2013}.
The photon detection efficiency was measured by preparing the ion into the $^2\mathrm{D}_{3/2} (F=1)$ state and measuring the 468~nm photons while repumping with 1079~nm light, following \mbox{Ramm \textit{et al.}~\cite{Ramm2013}}.
Our \ion{Ra}{225}\ hyperfine SPAM result compares favorably to recent results in similar systems~\cite{Christensen2020, Crain2019}.
The fidelity can be improved with a larger numerical aperture collection objective~\cite{Crain2019} or by shelving population in the $^2\mathrm{D}_{5/2}$ state with 728~nm laser light.

The \mbox{\gstransition} hyperfine transition is sensitive to the second-order Zeeman shift 
\mbox{$\Delta\nu_\mathrm{QZ} = C_2 B^2 \ ,$}
where $C_2$\ is the quadratic Zeeman shift coefficient and $B$\ is the magnitude of the magnetic field at the ion.
We measured $\Delta\nu_\mathrm{QZ}$ for different magnetic field strengths that were varied by moving a permanent magnet.
The magnet field is approximately perpendicular to the trap axis, see Fig.~\ref{fig:hfs_diagram}.
The field direction varies slightly when changing the magnet position and affects the relative coupling strength of the microwave $\pi$ and $\sigma^\pm$ transitions.
However, the line centers of the transitions are unaffected by these fluctuations and the linewidths are Fourier-limited.
We did not use shielding so the ambient B-field drifts over the course of a measurement.
The maximum observed drift rate is \systemnumber{\maxbfielddrift} and does not significantly affect our results.
From fitting these measurements we extract the quadratic Zeeman shift coefficient, \reportednumber{$C_2 =$ \qzcoeff}, see Fig.~\ref{fig:qz-shift}.
Our result is consistent with the calculated value of \systemnumber{\qzcoefftheory}~\cite{Versolato2011b}.

We perform interleaved Ramsey and Rabi spectroscopy sequences to measure the ground state hyperfine splitting.
First, we prepare the \mbox{$^2\mathrm{S}_{1/2} (F=0, m_F=0)$}\ state.
Then we scan either two \systemnumber{\SI{1.8}{\milli\second}} $\pi / 2$\ microwave pulses separated by \systemnumber{\SI{10}{\milli\second}} or a single \systemnumber{\SI{10}{\ms}} microwave pulse across the \mbox{$\mathrm{S}_{1/2} (F=0, m_F=0) \leftrightarrow~\mathrm{S}_{1/2} (F=1, m_F=0)$} $\pi$~transition.
After the Ramsey or Rabi measurement, we perform state detection.
To obtain the unperturbed $\pi$ transition frequency, \reportednumber{$\Delta\nu_\mathrm{QZ}$}\ \systemnumber{\mbox{($B=$ \gsbfield$)$}} is subtracted from the scanned frequency.
We average \systemnumber{\numramseys}~Ramsey and \systemnumber{\numrabis} interleaved Rabi measurements to determine the hyperfine splitting frequency, with a single representative Ramsey measurement shown in Fig.~\ref{fig:ramsey-scan}. 
This results in the ground state hyperfine constant \mbox{$A\left(\mathrm{S}_{1/2}\right) =$\reportednumber{\gsAhfs}.}
The statistical uncertainty is \reportednumber{\gsstaterr}, which is negligible compared to the \reportednumber{\gssyserr} total systematic uncertainty.

The dominant source of uncertainty of the $^2\mathrm{S}_{1/2}$\ hyperfine splitting comes from the second-order Zeeman shift, which shifts the frequency of the measured transition by $\Delta\nu_\mathrm{QZ}(B)$.
We measure $B$\ from the Zeeman splitting of the \mbox{$(F=0, m_F=0) \leftrightarrow (F=1, m_F=\pm1)$,} $\sigma^\pm$\ transitions, see Fig.~\ref{fig:qz-shift} inset.
The average frequency of the $\sigma^\pm$\ transitions is subtracted from the $\pi$ transition frequency to determine $\Delta\nu_\mathrm{QZ}(B)$.
The uncertainties for the quadratic Zeeman shift data points are derived from fits to the transitions, and contribute almost the entirety of the \reportednumber{\gsqzerr} to the measured ground state hyperfine splitting.

The second largest contribution to the systematic uncertainty is due to the ac~magnetic field $B_\mathrm{ac}$\ generated by imbalanced rf currents in the trap electrodes.
We measured the $\pi$~transition frequency for different trap rf powers to characterize this effect.
We found that the shift is consistent with zero, with $B_\mathrm{ac} \leq$\ \mbox{\reportednumber{\bacupperlimit}}, corresponding to a \reportednumber{\SI{0.5}{\Hz}} uncertainty.

The third largest contribution to the systematic uncertainty is inaccuracy of the rf source, which is locked to a GPS-disciplined reference (SRS~FS740).
We assign an uncertainty of \reportednumber{\gsrferr} to this systematic, derived from the reference \systemnumber{\SI{1e-11}{}} short-term stability.

We tested for ac~Stark shifts from probe laser light leakage but did not observe any frequency shifts due to this effect.
We also are not sensitive to shifts due to blackbody radiation, which we estimate to be below \systemnumber{\bbrupperlimit}.

\begin{figure}
    \includegraphics[width=\columnwidth]{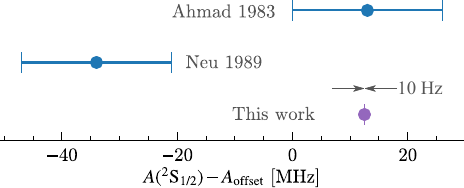}
    \captionsetup{justification=myjust,singlelinecheck=false}
    \caption{
    A comparison of our result (purple) with previous measurements ~\cite{Ahmad1983, Neu1989} (blue) of the $A(\mathrm{S}_{1/2})$\ hyperfine constant.  The plotted values are offset by $A_\mathrm{offset} = $\SI{-27.704}{\GHz}.
    }\label{fig:gs-hfs-compare}
\end{figure}

The $^2\mathrm{S}_{1/2}$ and $^2\mathrm{P}_{1/2}$ hyperfine splittings of \ion{Ra}{225} were first measured by laser spectroscopy of a hot atomic beam by \mbox{Ahmad \textit{et al.}~\cite{Ahmad1983}}.
The ground state hyperfine splitting was re-measured by \mbox{Neu \textit{et al.}~\cite{Neu1989}} and resulted in a  \mbox{\systemnumber{\groundstatediscrepancyMHz \ (\groundstatediscrepancysigma$\sigma$)}} discrepancy with the original measurement.
Of the two previous $A\left(\mathrm{S}_{1/2}\right)$ hyperfine constant measurements, our result is in agreement with the first measurement~\cite{Ahmad1983}, see Fig.~\ref{fig:gs-hfs-compare}.

We  measured the hyperfine splittings of the $^2\mathrm{D}_{3/2}$\ and $^2\mathrm{P}_{1/2}$\ excited states with laser spectroscopy. 
For both excited states, we scan over a transition with the first-order sideband of a fiber EOM.
The 468~nm EOM sideband drives the \mbox{$\mathrm{S}_{1/2} (F=1) \leftrightarrow~\mathrm{P}_{1/2} (F=1)$} transition and the 1079~nm EOM sideband drives the \mbox{$\mathrm{D}_{3/2} (F=1) \leftrightarrow~\mathrm{P}_{1/2} (F=1)$}\ transition.
We reference the laser carrier frequency by performing spectroscopy on the \mbox{$\mathrm{D}_{3/2} (F=2) \leftrightarrow \mathrm{P}_{1/2} (F=1)$} and the \mbox{$\mathrm{S}_{1/2} (F=1) \leftrightarrow \mathrm{P}_{1/2} (F=0)  $} transitions for the $^2\mathrm{D}_{3/2}$ and $^2\mathrm{P}_{1/2}$ hyperfine splitting measurements, respectively.

\begin{figure}
    \includegraphics[]{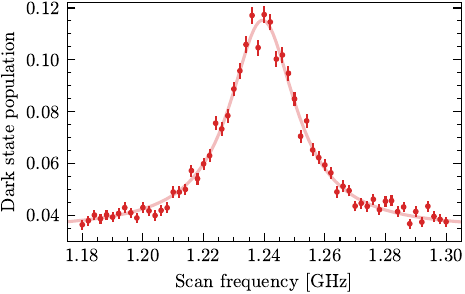}
    \captionsetup{justification=myjust,singlelinecheck=false}
    \caption{
    The $^2\mathrm{S}_{1/2} (F=0)$\ and \mbox{$^2\mathrm{D}_{3/2} (F=2)$}\ dark state population as the 1079~nm EOM first-order sideband is scanned over the \mbox{$\mathrm{D}_{3/2}~(F=1) \leftrightarrow \mathrm{P}_{1/2}~(F=1)$} transition. A Lorentzian fit determines the $^2\mathrm{D}_{3/2}$ hyperfine splitting.
    }
    \label{fig:d_3ov2_hfs}
\end{figure}

To measure the hyperfine splitting of the \mbox{$^2\mathrm{D}_{3/2}$}\ state, we prepare the population in the \mbox{$^2\mathrm{D}_{3/2} (F=1)$}\ hyperfine level, scan the 1079~nm EOM sideband frequency, and perform state detection.
The dark state population is maximized when the EOM sideband frequency is resonant with the $^2\mathrm{D}_{3/2}$\ hyperfine splitting, see Fig.~\ref{fig:d_3ov2_hfs}.
The hyperfine splitting is then determined from the difference of the \mbox{$\mathrm{D}_{3/2} (F=2) \leftrightarrow~\mathrm{P}_{1/2} (F=1)$}\ and \mbox{$\mathrm{D}_{3/2} (F=1) \leftrightarrow~\mathrm{P}_{1/2} (F=1)$} transition frequencies. 
The measured hyperfine splitting is $\reportednumber{\dsplitting}$, which corresponds to a hyperfine constant of \mbox{$A(\mathrm{D}_{3/2}) =$} \reportednumber{\dsAhfs}.
The statistical and systematic uncertainties are \reportednumber{\dsstaterr} and \reportednumber{\dssyserr}.

To measure the $^2\mathrm{P}_{1/2}$\ hyperfine splitting, we prepare the \mbox{$^2\mathrm{S}_{1/2} (F=1)$} bright state, scan the 468~nm EOM sideband frequency, and perform state detection.
The hyperfine splitting is determined from the frequency difference of the \mbox{$\mathrm{S}_{1/2} (F=1) \leftrightarrow~\mathrm{P}_{1/2} (F=0)$}\ and \mbox{$\mathrm{S}_{1/2} (F=1) \leftrightarrow~\mathrm{P}_{1/2} (F=1)$} transitions. 
The measured hyperfine splitting constant is \mbox{$A(\mathrm{P}_{1/2}) = $ \reportednumber{\psAhfs}}.
The uncertainty is dominated by the \reportednumber{\pssyserr} systematic uncertainty, primarily due to drift in the cavity that stabilizes the 468 nm laser. The statistical uncertainty is \reportednumber{\psstaterr}.
Our result is consistent with the previous measurement~\cite{Ahmad1983} and improves the precision by a factor of~\systemnumber{\mbox{1.9}}.

The $^2\mathrm{P}_{1/2}$\ and $^2\mathrm{D}_{3/2}$ hyperfine splitting measurements share the same sources of systematic uncertainty.
The leading source of uncertainty for both the $^2\mathrm{P}_{1/2}$\ and $^2\mathrm{D}_{3/2}$\ hyperfine measurements is drift of our laser stabilization cavities, which results in laser frequency drifts during measurements.
We characterize the drift and assign uncertainties of \reportednumber{\psdrifterr} to the $^2\mathrm{P}_{1/2}$ and \reportednumber{\dsdrifterr} to the $^2\mathrm{D}_{3/2}$\ measurements.

The combination of Zeeman splitting with imperfect linear laser polarization or with an unequal population distribution in the Zeeman sublevels used for state preparation results in a systematic shift of the $^2\mathrm{P}_{1/2}$\ and $^2\mathrm{D}_{3/2}$\ hyperfine frequencies.  For this shift we assign an uncertainty that corresponds to the largest Zeeman splitting of each transition, measured at a magnetic field of \mbox{\systemnumber{\esbfield}}.
The resulting uncertainty is \reportednumber{\pszeemanerr} for the  $^2\mathrm{P}_{1/2}$\ measurement, and \reportednumber{\dszeemanerr} for the $^2\mathrm{D}_{3/2}$\ measurement.

We investigated frequency shifts of the $^2\mathrm{P}_{1/2}$\ and $^2\mathrm{D}_{3/2}$\ states due to ac Stark effect from the probe lasers by varying the 468~nm laser power from \systemnumber{\pstarkpowers} and varying the 1079~nm laser power from \systemnumber{\dstarkpowers}.
We did not observe  ac~Stark shifts for either the $^2\mathrm{P}_{1/2}$\ or $^2\mathrm{D}_{3/2}$\ measurements, which is consistent with our calculated upper bound of \reportednumber{\SI{10}{\kHz}}.

Hyperfine measurements allow us to indirectly study nuclear magnetization, whose contribution to the hyperfine splitting is known as the Bohr-Weisskopf (BW) effect.  Direct nuclear calculations of the BW effect are challenging, with a \systemnumber{\directnuclearcalculationpct}\ uncertainty for the $^2$S$_{1/2}$ state of $^{225}$Ra$^+$~\cite{Ginges2017}. But, the effect can also be determined with a combination of atomic theory and hyperfine measurements~\cite{Ginges2018}.  The uncertainty from atomic theory (\systemnumber{\skipnikovprecision} in Skripnikov~\cite{Skripnikov2020} and \systemnumber{\cservenyprecision} in Cserveny \& Roberts~\cite{Cserveny2025}) is larger than the previous measurement uncertainty~\cite{Neu1989}.  Nevertheless our measurement combined with atomic theory shifts the BW value by several percent and reduces the uncertainty by a few tenths of a percent.  As theory improves, this hyperfine measurement will be increasingly important for our understanding of nuclear physics effects.  The ground state hyperfine measurement may also be used to improve knowledge of the BW effect for $^{225}$Ra$^+$ excited states~\cite{Ginges2018}.

In summary, we have operated a \ion{Ra}{225} ion trap for over \systemnumber{two years} in a sealed vacuum system and measured the hyperfine splitting of the states used for laser cooling and state detection.
The demonstrated long-term source of short-lived radium isotopes enables the development of a \ion{Ra}{225} ion optical clock, whose hyperfine structure is amenable to operation as a two-laser infrared clock~\cite{Holliman2023}.
Our SPAM result using only the three lowest electronic levels, i.e. there is no additional auxiliary shelving, and the ion's large hyperfine splitting make a case for \ion{Ra}{225} as a system for QIS applications~\cite{Zhang2025}.

Tests of time-reversal symmetry with \iso{Ra}{225} benefit from its nuclear structure, which enhances experimental sensitivity to the Schiff moment~\cite{Auerbach1996, Dobaczewski2005, Bishof2016}.
Sensitivity can be further enhanced by incorporating \iso{Ra}{225} into a molecule~\cite{Yu2021}, which can be synthesized with laser cooled \ion{Ra}{225}~\cite{Fan2021}.
The hyperfine and optical qubit features of \ion{Ra}{225} can be utilized in molecular spectroscopy techniques such as quantum logic spectroscopy~\cite{Schmidt2005, Chou2017}.
Finally, \ion{Ra}{225} ions are a potential route to trapped actinium ions, as \iso{Ra}{225} beta-decays to \ion{Ac}{225}, and therefore \ion{Ra}{225} decays could result in trapped $^{225}$Ac$^{2+}$, which supports laser cooling and state detection in two independent manifolds.  

\textit{Data availability.} The main data supporting the findings of this Letter are openly available \cite{Zenodo2026}.

We thank Sam Brewer, Wes Campbell, Christian Sanner, Qiming Wu, and Anthony Ransford for helpful discussions.
H.L. was supported by ONR Grant No.~N00014-21-1-2597 and M.F. was supported by DOE Award No. DE-SC0022034.
R.A.R., S.K., R.K., A.S., H.D., C.A.H., X.S., L.S.W., and A.M.J. were supported by the Heising-Simons Foundation Award No.~2022-4066, the W.M. Keck Foundation, NIST Award No.~60NANB21D185, NSF NRT Award No.~2152201, the Eddleman Center, the Noyce Initiative, NSF Award Nos.~2326810, 2146555, and 1912665, and ONR Award No. N00014-25-1-2131.
A.N.G. acknowledges the support of startup funds and Michigan State University.  The isotope used in this research was supplied by the U.S. Department of Energy Isotope Program, managed by the Office of Isotope R\&D and Production.


%

\end{document}